\documentclass[sigconf]{acmart}

\AtBeginDocument{%
  \providecommand\BibTeX{{%
    \normalfont B\kern-0.5em{\scshape i\kern-0.25em b}\kern-0.8em\TeX}}}

\usepackage{url}
\usepackage{xspace}
\usepackage{tabularx}
\usepackage{multirow}
\usepackage{graphics}
\usepackage{amsmath}
\usepackage{bm}
\usepackage{enumitem}
\usepackage{subfig}

\newcommand{\eg}{\emph{e.g.}\xspace}
\newcommand{\ie}{\emph{i.e.}\xspace}

\newcommand{\plus}{\scalebox{0.85}{$+$}}
\newcommand{\NA}{---}

\begin{document}

\settopmatter{printacmref=false} % Removes citation information below abstract
\renewcommand\footnotetextcopyrightpermission[1]{} % removes footnote with conference information in first column
\pagestyle{plain} % removes running headers

\title{Diffusion-based Time Series Data Imputation for Microsoft 365}

\author{
    Fangkai Yang$^*$, 
    Wenjie Yin$^\dagger$, 
    Lu Wang$^*$, 
    Tianci Li$^*$,
    Pu Zhao$^*$, 
    Bo Liu$^\mathsection$,
    Paul Wang$^\mathsection$,
    Bo Qiao$^*$,  
}
\author{
    Yudong Liu$^*$,
    % M\a arten Bj\"orkman$^\dagger$, 
    Mårten Björkman$^\dagger$, 
    Saravan Rajmohan$^\mathsection$,
    Qingwei Lin$^*$, 
    Dongmei Zhang$^*$
}
\affiliation{
    \institution{
        Microsoft Research$^*$\country{} \quad
        Microsoft 365$^\mathsection$\country{} \quad
        KTH Royal Institute of Technology$^\dagger$
    }
}

\renewcommand{\shortauthors}{Fangkai Yang et al.}
 
\begin{abstract}
Reliability is extremely important for large-scale cloud systems like Microsoft 365. Cloud failures such as disk failure, node failure, etc. threaten service reliability, resulting in online service interruptions and economic loss. Existing works focus on predicting cloud failures and proactively taking action before failures happen. However, they suffer from poor data quality like data missing in model training and prediction, which limits the performance. In this paper, we focus on enhancing data quality through data imputation by the proposed Diffusion\textsuperscript{\plus}, a sample-efficient diffusion model, to impute the missing data efficiently based on the observed data. Our experiments and application practice show that our model contributes to improving the performance of the downstream failure prediction task.

\end{abstract}




\maketitle

\section{Introduction}
% require high service reliability
Microsoft 365 cloud platform is a large-scale online service system and serves millions of customer workloads on a 24/7 basis. It is extremely critical to ensure high reliability as any cloud failure will result in financial loss and degradation of user experience~\cite{jayathilaka2017performance,chen2019outage,levy2020predictive}. 
% cloud failure prediction is a critical issue
% disk, node, memory failure prediction 
However, cloud failure, including hardware failure and software failure, is inevitable in large-scale systems~\cite{botezatu2016predicting,meza2015large,gao2020task}.
Recent research and works~\cite{chen2019outage,luo2021ntam,liu2022multi,ma2022empirical} have proposed approaches to predict cloud failures before they actually happen and take actions proactively to mitigate potential failures, thus minimizing the negative impact of cloud failure. 
% the necessity of improving data quality
Although significant progress has achieved good results in practice, these failure prediction methods still suffer from the issues of data missing~\cite{deb2017aesop, chai2020human, luo2021ntam}. 
% the cause of data missing
Data missing is a practical and ubiquitous problem in large cloud systems caused by data delay~\cite{lu2020making}, monitoring error~\cite{ward2014observing}, etc. In this paper, rather than designing better failure prediction models, we focus on a new perspective of enhancing the data quality by imputing missing data to improve the performance of downstream cloud failure prediction. 

\begin{figure}[t]
  \centering
  \includegraphics[width=0.95\linewidth]{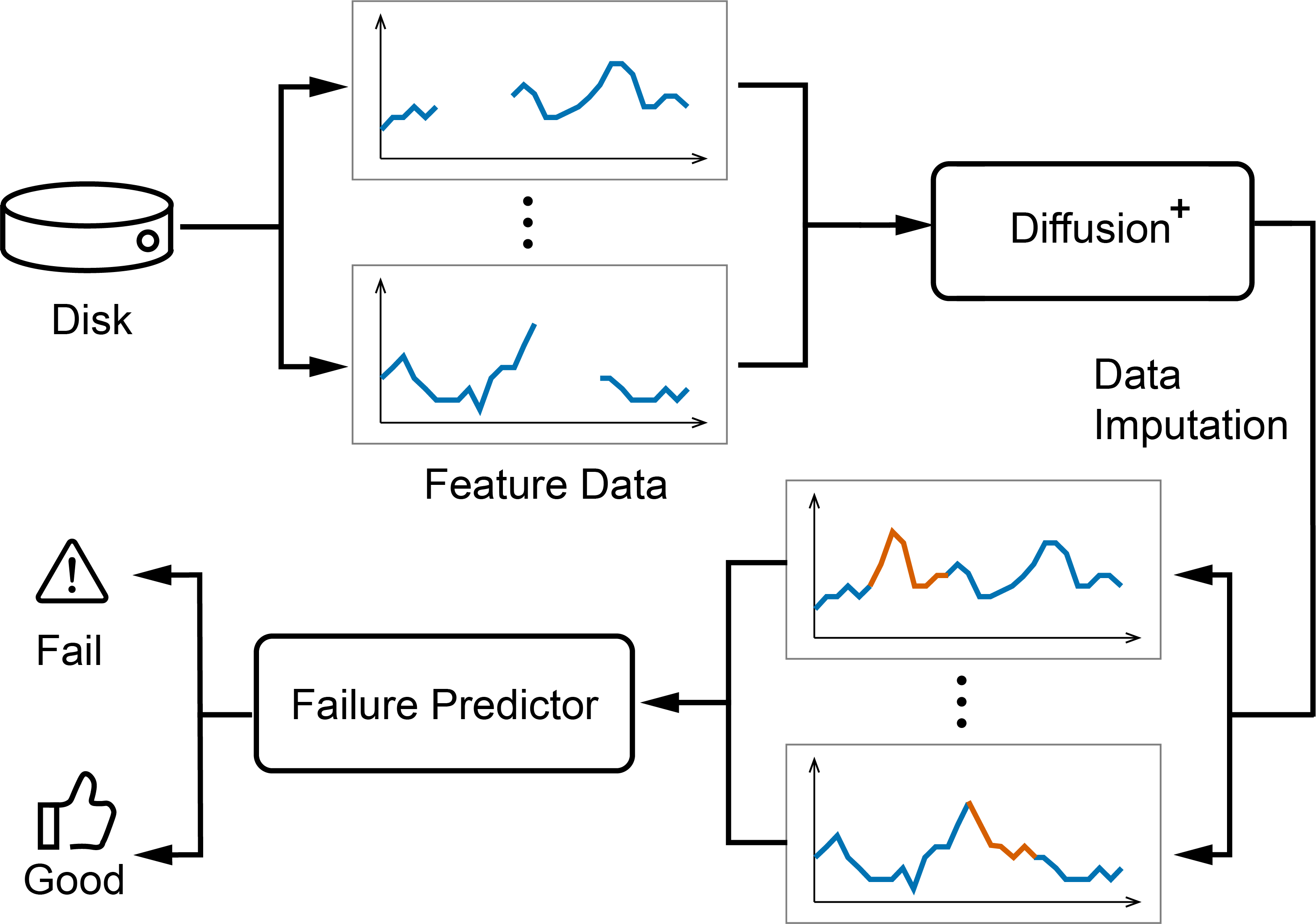}
  \caption{The overview of data imputation with downstream failure prediction tasks.}
  \label{fig:imputationprocess}
  \vspace{-3mm}
%   \vspace{\vspacereduce}
\end{figure}

There exist a large number of studies of data imputation concerning images and time series data~\cite{pratama2016review,cao2018brits,fortuin2020gp,tashiro2021csdi}. However, very few focus on time series data imputation in the domain of cloud systems, and rules-based and statistical approaches are commonly used in industry~\cite{ma2022empirical}. Most importantly, there lacks an end-to-end evaluation of data imputation with the downstream tasks, and the effect of different data imputation methods is still unexplored connecting to the downstream tasks that utilize the data. In this paper, we leverage the success of the diffusion models~\cite{sohl2015deep,ho2020denoising}, which have outperformed state-of-the-art generative models with higher sample quality, and we propose a new diffusion model, \ie, Diffusion\textsuperscript{\plus}, to impute missing data with high efficiency. Figure~\ref{fig:imputationprocess} shows the overview of the process. We use the diffusion model to do data imputation, and the imputed data is fed into the downstream failure prediction task for model training and prediction. We select disk failure prediction as the downstream task since disk failure is one of the most frequent failures in cloud systems~\cite{sankar2013datacenter,botezatu2016predicting}, and our model can be easily adapted to other downstream tasks in cloud scenarios. Moreover, the slow sampling issue of the diffusion model restricts its application in industry. Inspired by the most recent work~\cite{lu2022dpm}, we improve the diffusion sampling efficiency with at least 4$\times$ speed up without degrading the downstream prediction task.

Our main contributions are summarized as follows:
\begin{itemize}[noitemsep,topsep=0pt]
    \item We propose a new perspective of improving cloud failure prediction by imputing missing data.
    \item Inspired by the diffusion model, we propose a new diffusion model with better imputation performance and higher sampling efficiency in the cloud scenario.
    \item We conduct extensive experiments on industrial data and demonstrate that our model improves the performance on the downstream task.
\end{itemize}

% \section{Challenges and motivation}

\section{Methodology}

\subsection{Problem Formulation}

In practice, a disk's status vector is recorded at each timestamp (\eg, hourly), which is a multivariate time series data $x\in \mathbb{R}^{K\times L}$ where $K$ is the number of features and $L$ is the length of time series. At each timestamp $l$, some data features are missing in $x_l$, then we can partition $x_l$ into the missing part $x^{ms}_l=\{x_l^k | x_l^k \; \text{is missing} \}^{1:K}$ and the observed part $x^{ob}_l=\{x_l^k | x_l^k \; \text{is observed} \}^{1:K}$, \ie, $x_l=x^{ms}_l \cup x^{ob}_l$. Our goal is to do data imputation for $x^{ms} = \{x^{ms}_l\}_{1:L}$ given $x^{ob}=\{x^{ob}_l\}_{1:L}$ for all status feature vectors $x$, and the imputed status feature vectors are then fed toward downstream prediction tasks, which is trained to predict whether a disk will fail or not. 

\subsection{Overview of Diffusion\textsuperscript{\plus} Model}\label{sec:overview}
Denoising diffusion probabilistic models (DDPM)~\cite{sohl2015deep,ho2020denoising}, known as diffusion models (DM) for brevity, are a class of generative models inspired by non-equilibrium thermodynamics. DMs consist of a forward process and a reverse process. 
In the forward process, DMs define a fixed Markov chain of $T$ diffusion steps to slowly add noise to the data $x_0\in \mathbb{R}^{K\times L}$ until the data distribution is close to a standard Gaussian distribution $x_T\in \mathbb{R}^{K\times L}$. Note that the subscripts in $x_0$ and $x_T$ represent the diffusion step, \ie, $x_0=\{x_{0,l}\}_{1:L}$, and we omit $l$ for simplicity. 
The forward process is defined as:
\begin{equation}
\begin{aligned}
    q(x_{1:T}|x_0) &=  \prod_{t=1}^{T}q(x_{t}|x_{t-1}),\\
    q(x_{t}|x_{t-1})&=  \mathcal{N}(x_{t};\sqrt{1-\beta_{t}}x_{t-1}, \beta_{t}\mathbf{I}),
\end{aligned}
\end{equation}
where $\beta_{1},\beta_{2}, \cdots, \beta_{T}$ are the fixed noise schedulers for controlling the noise scale~\cite{ho2020denoising}. 
 
\begin{figure}[t]
  \centering
  \includegraphics[width=0.98\linewidth]{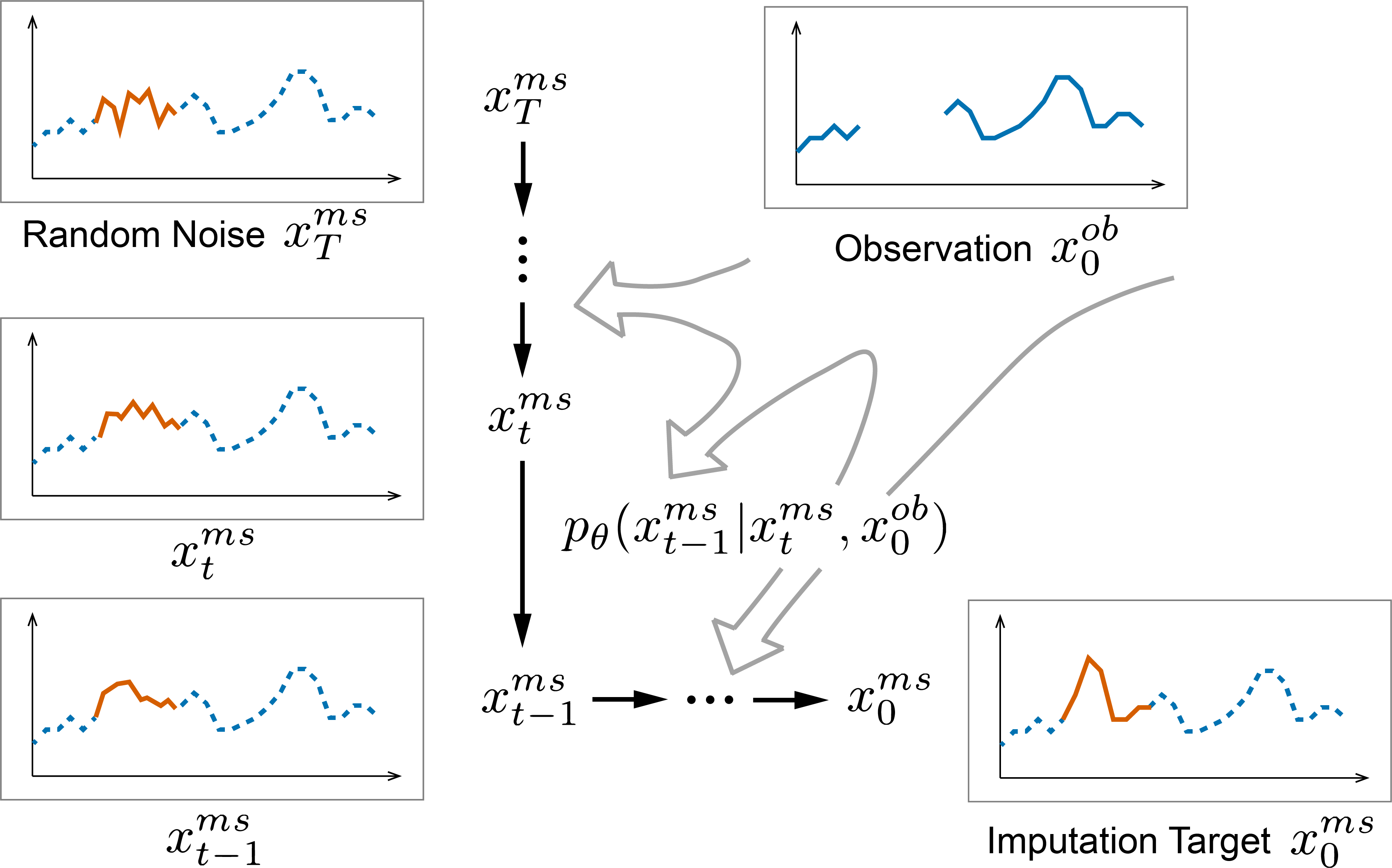}
  \caption{Data imputation with the reverse process of the diffusion model.}
  \label{fig:diffusionmodel}
  \vspace{-3mm}
%   \vspace{\vspacereduce}
\end{figure}

On the other hand, in the reverse process, DMs learn to reverse the forward process by denoising to get the desired data distribution from the noise distribution, \ie, sampling from $q(x_{t-1}|x_{t})$ will be able to create the true sample $x_0$ from a Gaussian noise $x_T$. However, it is non-trivial to estimate $q(x_{t-1}|x_{t})$, and we learn to model $p_{\theta}(\cdot)$ as the approximate estimation.
We adopt the conditional diffusion model \cite{tashiro2021csdi} which uses the observation $x_0^{ob}$ as the condition to generate imputation targets $x_0^{ms}$. More specifically, the goal of data imputation is to estimate the true conditional data distribution $q(x_0^{ms}|x_0^{ob})$ with a model distribution $p_{\theta}(x_0^{ms}|x_0^{ob})$, and the missing data $x_0^{ms}$ can be sampled from $p_{\theta}(\cdot)$ as shown in Figure~\ref{fig:diffusionmodel}. We model $p_{\theta}(x_0^{ms}|x_0^{ob})$ with the diffusion model in the reverse process:

{\small
\begin{equation}\label{eqn:diffusionmodel}
\begin{aligned}
    p_{\theta}(x_{0:T}^{ms}|x_0^{ob}) = p(x_T^{ms})\prod_{t=1}^{T}p_{\theta}(x_{t-1}^{ms} | x_{t}^{ms}, x_0^{ob}), \ x_T^{ms}\sim \mathcal{N}({\bm{0,I}}), \\
    p_{\theta}(x_{t-1}^{ms} | x_{t}^{ms}, x_0^{ob})=\mathcal{N}(x_{t-1}^{ms};\bm{\mu}_{\theta}(x^{ms}_t,t|x_0^{ob}),\sigma_{\theta}(x^{ms}_t,t|x_0^{ob})\bm{I})
\end{aligned}
\end{equation}
}

% where $\bm{\mu}_{\theta}(\cdot)$ and $\sigma_{\theta}(\cdot)$ are parameterized following the denoising diffusion probabilistic models (DDPM) \cite{ho2020denoising}.

We define a conditional denoising function $\epsilon_{\theta}$ in the reverse process to estimate $\bm{\mu}_{\theta}(\cdot)$ and $\sigma_{\theta}(\cdot)$ of the distribution $p_{\theta}(x_{t-1}^{ms} | x_{t}^{ms}, x_0^{ob})$. In particular, $\bm{\mu}_{\theta}(x^{ms}_t,t|x_0^{ob})=\bm{\mu}^{DDPM}(x^{ms}_t,t,\bm{\epsilon}_{\theta}(x^{ms}_t,t|x_0^{ob}))$ and $\sigma_{\theta}(x^{ms}_t,t|x_0^{ob})=\sigma^{DDPM}(x^{ms}_t,t)$, where $\bm{\mu}^{DDPM}(\cdot)$ and $\sigma^{DDPM}(\cdot)$ are the parameterization functions in 
denoising diffusion probabilistic models (DDPM)~\cite{ho2020denoising}. Then, given $\bm{\epsilon}_{\theta}$ and $x_0^{ob}$, we can sample $x_0^{ms}$ in the reverse process in Equation~\ref{eqn:diffusionmodel}, where $\bm{\epsilon}_{\theta}$ is trainable. 

\noindent{\textbf{Training}}.
Since we do not have the ground-truth missing values, we first do zero imputation for the missing data, and then we randomly partition the observation $x_0^{ob}$ into two parts: the conditional observation $\hat{x}_0^{ob}$, and the masked target that needs imputation $\hat{x}_0^{ms}$. 
Our model is then trained in a self-supervised learning manner~\cite{devlin2018bert} to do data imputation for $\hat{x}_0^{ms}$ given $\hat{x}_0^{ob}$, and the imputation performance is evaluated on $\hat{x}_0^{ms}$.
With the formulated forward process and the reverse process, the training process optimizes the log-likelihood in the reverse process by maximizing the variational lower bound. 
The training is performed for all diffusion steps. and it is trained by minimizing the simplified objective function:

\begin{equation}
    \min_{\theta} \mathcal{L} = \mathbb{E}_{\bm{\epsilon}\sim \mathcal{N}(\bm{0},\bm{I})} ||\bm{\epsilon}-\bm{\epsilon}_{\theta}(\hat{x}^{ms}_t,t|\hat{x}_0^{ob}) ||_2^2
\end{equation}.

\noindent{\textbf{Inference}}.
When the training is done, we have good modeling of $p_{\theta}(\cdot)$.
Given the real observation $x_0^{ob}$ as the conditional observation,  we could impute the missing data $x_0^{ms}$ with the reverse generation process $x_{t-1}^{ms} \sim p_\theta(x_{t}^{ms}|x_0^{ob})$ according to Equation~\ref{eqn:diffusionmodel}.
 
For each sample with missing data, we generate 100 data imputations and take their median as the final imputed results. 
The data imputation is conducted over the whole dataset before training the downstream failure prediction models.

\noindent{\textbf{Speed up}}.
As shown in Figure~\ref{fig:diffusionmodel}, DMs suffer from slow sampling as they require a large number of diffusion steps $T$ of running large neural networks to draw one sample~\cite{lu2022dpm}, which makes it inefficient and impractical for data imputation in industry and becomes a bottleneck for the downstream tasks. Inspired by recent work~\cite{lu2022dpm}, we speed up the data imputation by reducing the diffusion steps in the reverse process without any further training. 
The sampling of DMs in the reverse process can be viewed alternatively as solving corresponding ordinary differential equations (ODEs)~\cite{jolicoeur2021gotta,song2020score}, and the sampling process is done by ODE solvers~\cite{atkinson2011numerical,lu2022dpm} which results in high-quality and few-step sampling.
% Specifically, given an initial sample $x_{T}$, and $M+1$ fewer diffusion steps decreasing from $t_0=T$ to $t_M=0$, the sequence can be computed iteratively. The time steps $\{t_i\}_{i=0}^{M}$ need to be specified in advance and we adopt a uniform step size schedule. 
Specifically, the noise scheduler in $\bm{\mu}_{\theta}(\cdot)$ and $\sigma_{\theta}(\cdot)$ is updated by the ODE solver, and we adopt a uniform step size schedule to determine $M$ ($\ll T$) steps. Then the diffusion steps $T$ in the reverse process is reduced to $M$ steps.

\section{Experiments}

\subsection{Experimental Settings}
The data we used for experiments were collected from the Microsoft 365 online service system in recent 6 months. 
The data is in the SMART format (Self-Monitoring, Analysis and Reporting Technology)~\cite{allen2004monitoring}, which records the disk status and provides
important indicators during the lifetime of disks. 
We predict the disk failure based on 72-hour data. All experiments are performed on a workstation equipped with AMD EPYC 7V12 64-Core CPUs, NVIDIA Tesla T4 GPU with CUDA 10.1, and running Linux (16.04.5) OS.

\subsection{Baselines}
Following the previous work~\cite{fortuin2020gp,tashiro2021csdi,ma2022empirical}, we use imputation baselines as follows:

\noindent{\textbf{Zero imputation (Z)}}: Zero imputation replaces the missing data with zero, which is the most intuitive way.

\noindent{\textbf{Forward imputation (F)}}: Forward imputation~\cite{little2002single} is a single imputation method that replaces the missing data with the previously observed value.

\noindent{\textbf{Linear interpolation (L)}}: Linear interpolation~\cite{read1999linear} interpolates the missing data by linear curve fitting.

\noindent{\textbf{BRITS}}: BRITS~\cite{cao2018brits} is an RNN-based approach that utilizes a bi-directional recurrent neural network that handles the missing data considering the forward and backward temporal dependency.

\noindent{\textbf{Variational Autoencoders (VAE)}}: VAE \cite{kingma2013auto,fortuin2020gp} is a generative model that learns a probability distribution representing the data, and the missing data is sampled from the estimated distribution. 

Following the previous work on disk failure prediction~\cite{luo2021ntam,liu2022multi}, we use the downstream prediction baselines: long short-term memory (LSTM)~\cite{zhang2018layerwise}, Transformer (Trans)~\cite{luo2021ntam}, and temporal convolutional neural network (TCNN)~\cite{sun2019system}.

\subsection{Experimental Results}

In this section, we aim to address three research questions:

\begin{itemize}[noitemsep,topsep=0pt,leftmargin=*]
    \item \textbf{RQ1: Do the diffusion and Diffusion\textsuperscript{\plus} models impute missing data effectively?}
\end{itemize}

As mentioned in Section~\ref{sec:overview}, we randomly mask parts of the observation $x_0^{ob}$ as the imputation target $\hat{x}_0^{ms}$, and we train and evaluate data imputation models with 10\%, 50\%, 90\% masked missing ratio following previous work~\cite{tashiro2021csdi}. Note that Z, F, and L are rule-based methods without training, and we list them for reference.

We first present the quantitative results. We adopt two metrics following previous work~\cite{tashiro2021csdi} to evaluate the performance of data imputation, \ie, MAE (mean absolute error) and CRPS (continuous ranked probability score), where CRPS~\cite{matheson1976scoring} is usually used to measure the compatibility of an estimated probability distribution with an observation. For the deterministic imputation methods, \ie, Z, F, L, and BRITS, we only use MAE since they are not probabilistic imputation methods. As for probabilistic imputation methods (VAE, Diffusion, and Diffusion\textsuperscript{\plus}), we generate 100 samples for each missing data sample to estimate the probability distribution of the missing data with the metric CRPS. The MAE of the probabilistic imputation methods is computed using the median of 100 generated samples. Note that the data are normalized within each feature dimension in the evaluation. As shown in Table~\ref{tab:dataimputation}, the diffusion model has the lowest MAE, 49\%-95\% less compared with baselines. It suggests that the diffusion model is more effective in capturing the feature and temporal dependency. The diffusion model also shows the lowest CRPS metric compared with VAE, which indicates its capability of generating more realistic distributions. Diffusion\textsuperscript{\plus} model has a very close performance as the diffusion model, \ie, second best, in general (excluding models trained under 90\% missing ratio). Most imputation approaches have better performance with smaller missing ratios since more observations are available. 
Thus, we use the models trained under the 10\% missing ratio for imputation, \ie, models with the best performance trained with three missing ratios. 

\begin{table}[t]
\centering
\caption{Data imputation performance evaluated with MAE and CRPS (lower is better). CRPS is only available for probabilistic imputation methods.}\label{tab:dataimputation}
\resizebox{\columnwidth}{!}{
\begin{tabular}{ccccccc}
\toprule
\multirow{3}{*}{\textbf{Approach}} & \multicolumn{6}{c}{\textbf{Mssing Ratio (\%)}}                                                      \\ \cline{2-7} 
                                   & \multicolumn{2}{c}{\textbf{10}} & \multicolumn{2}{c}{\textbf{50}} & \multicolumn{2}{c}{\textbf{90}} \\ \cline{2-7} 
                                   & \textbf{MAE}   & \textbf{CRPS}  & \textbf{MAE}   & \textbf{CRPS}  & \textbf{MAE}   & \textbf{CRPS}  \\ \midrule
Z   &  0.429  &  \NA  &  0.428   &   \NA  &   0.429     &  \NA  \\
F  &  0.047   & \NA  &   0.057  & \NA  &  0.111   & \NA    \\
L  &  0.063  & \NA  &  0.064  & \NA  &  0.068  & \NA    \\
BRITS  & 0.052 &  \NA   &  0.054 &  \NA  &  0.081  &  \NA   \\
VAE  &  0.039    &  0.613     &   0.045     & 0.616  &    0.075    &     0.648   \\
Diffusion   &\textbf{0.020} & 0.049 & \textbf{0.021} & \textbf{0.040}& \textbf{0.053} &  \textbf{0.131} \\
Diffusion\textsuperscript{\plus}  & 0.025&  \textbf{0.046} &   0.034 &   0.068  & 0.099 & 0.253 \\ \bottomrule
\end{tabular}
}
\vspace{-3mm}
\end{table}

\begin{figure}[t]
  \centering
  \includegraphics[width=0.98\linewidth]{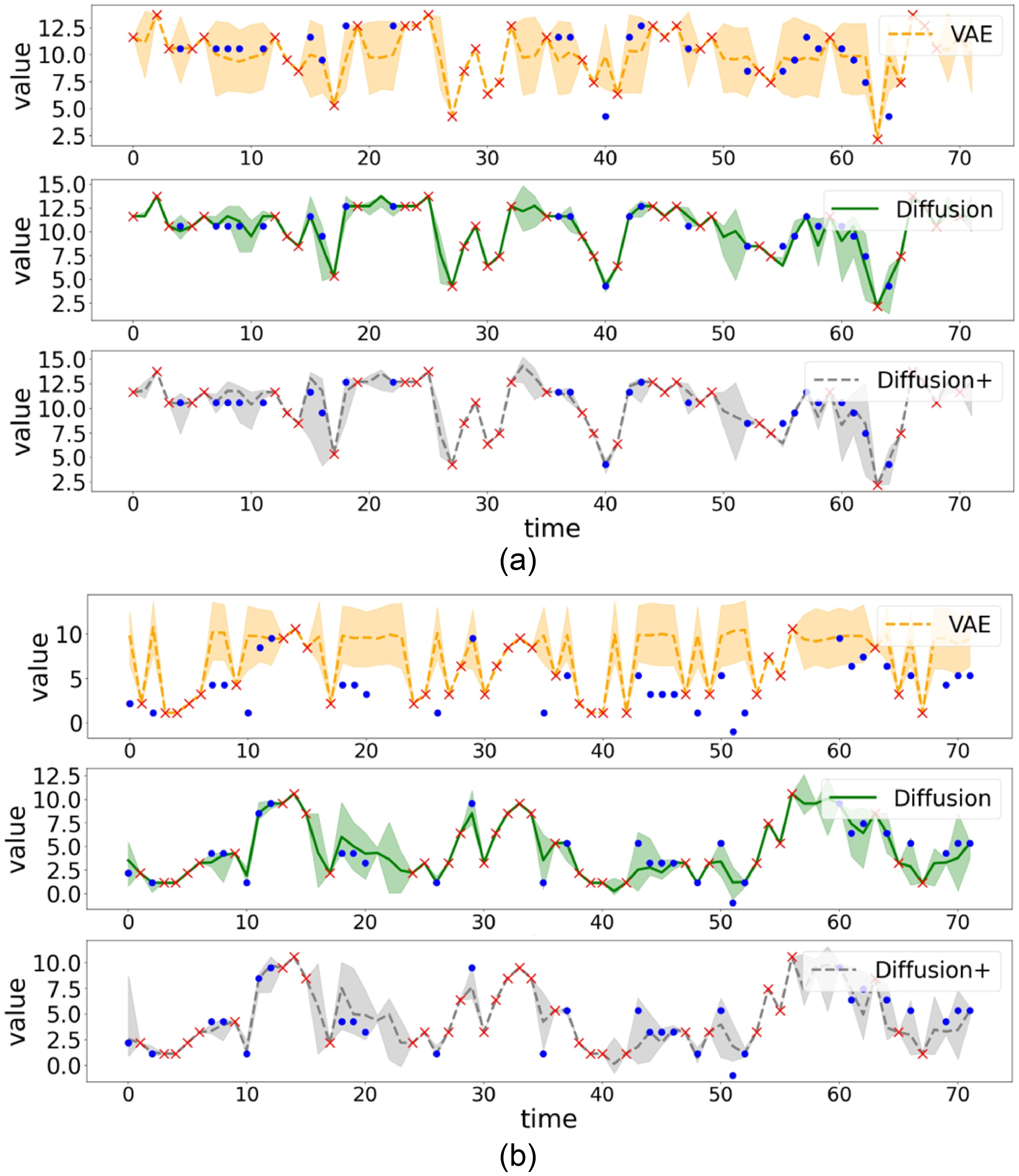}
\caption{Two data imputation examples of VAE, Diffusion, and Diffusion\textsuperscript{\plus}. Each example is a time series sample of one feature. The red crosses represent observed values and the blue dots represent the masked observation data for imputation targets. The shaded areas are 5\% and 95\% quantiles and the line is the median value of imputation.}
  \label{fig:imputationexample}
  \vspace{-5mm}
%   \vspace{\vspacereduce}
\end{figure}

We also provide imputation examples (shown in Figure~\ref{fig:imputationexample}). The diffusion and Diffusion\textsuperscript{\plus} models generate imputations with high confidence and the imputation distributions tightly cover masked missing targets (blue dots). VAE imputations have larger variations and cannot cover the missing targets. 

\begin{itemize}[noitemsep,topsep=0pt,leftmargin=*]
    \item \textbf{RQ2: Does data imputation contribute to improving the downstream disk failure prediction task?}
\end{itemize}
% \lu{it seems the higher imputation error does not result in higher failure prediction error}
We impute the ground-truth missing data in the entire dataset and feed the imputed data to downstream failure prediction tasks. As shown in Table~\ref{tab:prediction}, with all prediction methods, the diffusion imputation model achieves the best performance in precision and F1-score, and also in recall for most cases.  
In the domain of cloud failure prediction, F1-score is the most important metric~\cite{ma2022empirical, liu2022multi}. Diffusion\textsuperscript{\plus} shows a very close performance as the diffusion model in F1-score, and it demonstrates the second best of all the other approaches. Compared with different prediction methods, Trans achieves the best performance in F1-score. Note that we use advanced failure prediction models in practice~\cite{liu2022multi, luo2021ntam}, which have better prediction performance than the prediction baselines.

\begin{table}[t]
\centering
\caption{Failure prediction performance with different data imputation methods on three metrics, \ie, precision, recall, and F1-score.} \label{tab:prediction}
\begin{tabular}{cccc}
\toprule
\textbf{Approach} & \textbf{Precision}   & \textbf{Recall}      & \textbf{F1-score}    \\ \midrule
Z+LSTM            & 60.00  & 50.45   & 54.81                      \\
F+LSTM            & 64.69  &  46.41  &  54.05                     \\
L+LSTM            & 59.13  & 44.74 & 50.94                      \\
BRITS+LSTM          &  61.20  & 50.22   & 55.17        \\
VAE+LSTM          &  62.07  & 52.84 &  57.08  \\
Diffusion+LSTM    & \textbf{66.75} & \textbf{55.49}  & \textbf{60.60}          \\
Diffusion\textsuperscript{\plus}+LSTM    & 65.96   &54.23 & 59.52 \\ \hline
Z+Trans             & 62.84  & 51.57 & 56.65                      \\
F+Trans            & 68.15  & 47.98  & 56.32                      \\
L+Trans            & 62.87 & 48.21 & 54.57                      \\
BRITS+Trans          & 64.81  & 52.92  &  58.26         \\
VAE+Trans          &  66.85  &  52.45  &  58.78       \\
Diffusion+Trans    & \textbf{74.05}  & \textbf{53.59} & \textbf{62.18} \\
Diffusion\textsuperscript{\plus}+Trans    & 72.01& 52.34  & 60.62 \\ \hline
Z+TCNN            & 60.60  & 50.00  &  54.79                      \\
F+TCNN            & 61.05  &   50.66 &  55.37                     \\
L+TCNN            & 59.44  & 47.55 & 52.83                     \\
BRITS+TCNN          & 60.61  &  50.32  & 54.99        \\
VAE+TCNN          & 60.50   &  \textbf{54.64}   &  57.42        \\
Diffusion+TCNN    &\textbf{79.24}  & 48.31 & \textbf{60.03}  \\
Diffusion\textsuperscript{\plus}+TCNN    & 72.93  & 49.78 & 59.17 \\ 
\bottomrule
\end{tabular}
\end{table}

\begin{itemize}[noitemsep,topsep=0pt,leftmargin=*]
    \item \textbf{RQ3: Does our Diffusion\textsuperscript{\plus} model speed up the sampling process in diffusion models?}
\end{itemize}

\begin{figure}[t]
  \centering
  \includegraphics[width=0.95\linewidth]{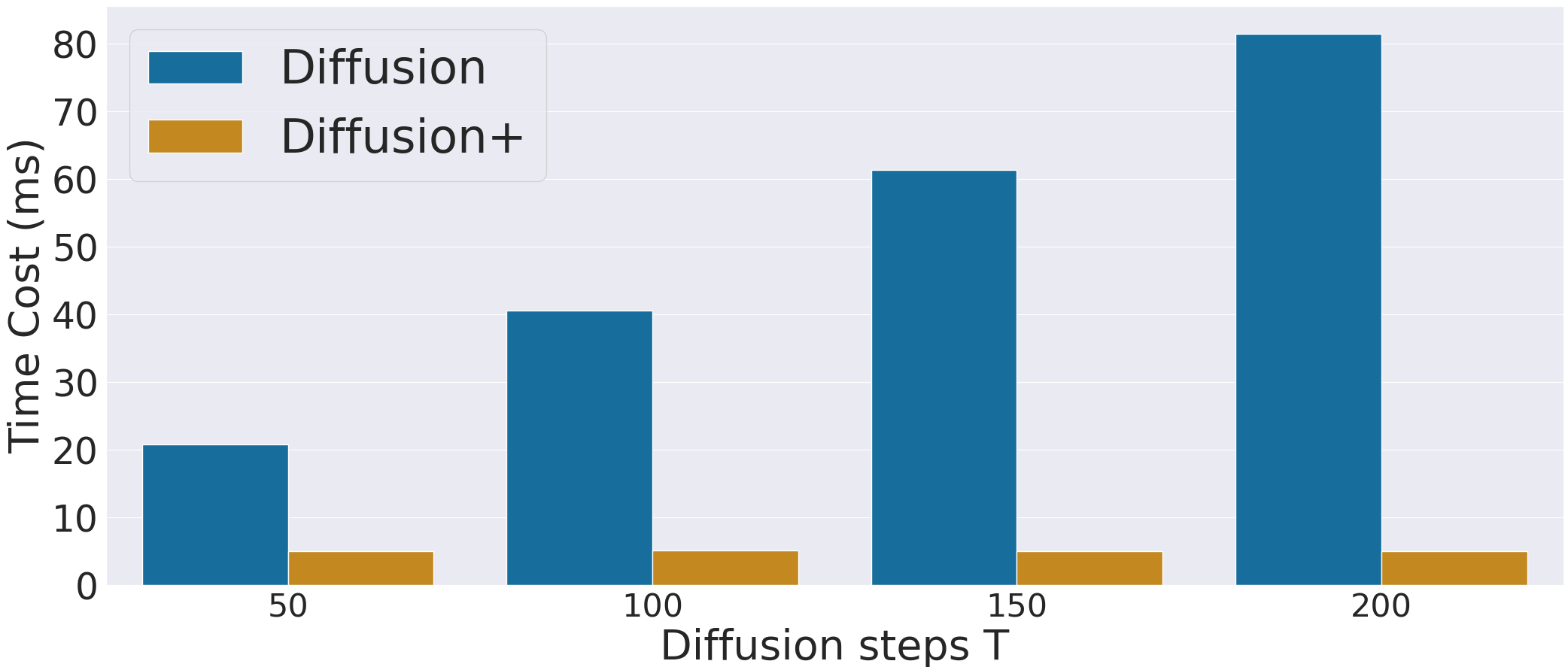}
  \caption{The time cost (ms) for each diffusion imputation.}
  \label{fig:timecost}
  \vspace{-3mm}
%   \vspace{\vspacereduce}
\end{figure}

Diffusion models suffer from slow sampling issues since generating one sample requires a large number of diffusion steps. Diffusion\textsuperscript{\plus} model aims to speed up the sampling process with only a few sampling steps without degrading the performance too much. As discussed in RQ1 and RQ2, Diffusion\textsuperscript{\plus} model achieves similar performance as the diffusion model. Then we conduct the analysis on time cost for imputing each sample. Figure~\ref{fig:timecost} shows the averaged time cost of data imputation for each data sample. As the diffusion step $T$ grows, the time cost for the diffusion model increases accordingly, while Diffusion\textsuperscript{\plus} has a stable time cost that needs far fewer diffusion steps and it has at least 4$\times$ speed up, and the speed up is more obvious with the increase of diffusion steps.

\section{Application in practice}
We have run our Diffusion\textsuperscript{\plus} model for one month on Microsoft 365, which contains millions of disks. In particular, our model takes effect in the data process phase of the current disk failure prediction pipeline~\cite{liu2022multi}. The SMART data is first collected by a data collection service, transferred by a distributed streaming tool, and stored in Azure. Then, our model imputes missing data before sending it to the feature engineering of the downstream prediction tasks. 
We conduct A/B testing to measure the effectiveness of our model and its contribution to service reliability. We monitor the reduction of virtual machine (VM) interruptions by taking proactive failure mitigation based on the prediction. Compared with the original data process phase without Diffusion\textsuperscript{\plus}, the interruptions have reduced the VM interruptions and enhanced the service reliability to avoid potential financial loss.

\section{Related work}
\noindent{\textbf{Cloud failure prediction.}}
There exist many studies on cloud failure prediction~\cite{botezatu2016predicting,meza2015large,gao2020task}, and they are commonly treated as binary classification problems~\cite{liu2022multi}. They use collected monitoring metrics from services in a time window to predict whether there will be a failure in the near future. They can capture temporal dependency to make a good prediction. However, missing data is a critical issue for these approaches since it requires the prediction models to infer missing information, and it usually results in poor prediction performance~\cite{fletcher2020missing,alcaraz2022diffusion}. Our paper is orthogonal to these failure prediction methods, and it offers a new perspective to improve cloud failure prediction by enhancing the data quality.

\noindent{\textbf{Time series data imputation.}}
Time series data imputation is a rich topic~\cite{little2014joys}. In particular, deep learning models including RNN-based approaches~\cite{cao2018brits,che2018recurrent,liu2019naomi} and generative models~\cite{fortuin2020gp,tashiro2021csdi,luo2018multivariate} can capture the temporal dependency of the time series and generate better data imputation than rule-based and statistical methods. Different from these methods, our paper not only evaluates the imputation quality but also focuses on the end-to-end performance of data imputation with practical industrial problems, \ie, disk failure prediction, and we speed up the diffusion-model-based data imputation to make it applicable in industry.

\section{Conclusion}
In this paper, we focus on enhancing the data quality for disk failure prediction by imputing missing data. We propose our Diffusion\textsuperscript{\plus} model based on diffusion models which imputes missing data effectively and efficiently. Our experiments on industrial datasets collected in Microsoft 365 and A/B testing show that our model outperforms baselines with fast sampling speed and contributes to enhancing the failure prediction tasks, and then improving the reliability of the Microsoft 365 cloud platform.

%%
%% The acknowledgments section is defined using the "acks" environment
%% (and NOT an unnumbered section). This ensures the proper
%% identification of the section in the article metadata, and the
%% consistent spelling of the heading.
% \begin{acks}
% To Robert, for the bagels and explaining CMYK and color spaces.
% \end{acks}

%%
%% The next two lines define the bibliography style to be used, and
%% the bibliography file.
\newpage
\bibliographystyle{ACM-Reference-Format}
\balance
\bibliography{www2023}

%%% -*-BibTeX-*-
%%% Do NOT edit. File created by BibTeX with style
%%% ACM-Reference-Format-Journals [18-Jan-2012].

\begin{thebibliography}{38}

%%% ====================================================================
%%% NOTE TO THE USER: you can override these defaults by providing
%%% customized versions of any of these macros before the \bibliography
%%% command.  Each of them MUST provide its own final punctuation,
%%% except for \shownote{}, \showDOI{}, and \showURL{}.  The latter two
%%% do not use final punctuation, in order to avoid confusing it with
%%% the Web address.
%%%
%%% To suppress output of a particular field, define its macro to expand
%%% to an empty string, or better, \unskip, like this:
%%%
%%% \newcommand{\showDOI}[1]{\unskip}   % LaTeX syntax
%%%
%%% \def \showDOI #1{\unskip}           % plain TeX syntax
%%%
%%% ====================================================================

\ifx \showCODEN    \undefined \def \showCODEN     #1{\unskip}     \fi
\ifx \showDOI      \undefined \def \showDOI       #1{#1}\fi
\ifx \showISBNx    \undefined \def \showISBNx     #1{\unskip}     \fi
\ifx \showISBNxiii \undefined \def \showISBNxiii  #1{\unskip}     \fi
\ifx \showISSN     \undefined \def \showISSN      #1{\unskip}     \fi
\ifx \showLCCN     \undefined \def \showLCCN      #1{\unskip}     \fi
\ifx \shownote     \undefined \def \shownote      #1{#1}          \fi
\ifx \showarticletitle \undefined \def \showarticletitle #1{#1}   \fi
\ifx \showURL      \undefined \def \showURL       {\relax}        \fi
% The following commands are used for tagged output and should be
% invisible to TeX
\providecommand\bibfield[2]{#2}
\providecommand\bibinfo[2]{#2}
\providecommand\natexlab[1]{#1}
\providecommand\showeprint[2][]{arXiv:#2}

\bibitem[Alcaraz and Strodthoff(2022)]%
        {alcaraz2022diffusion}
\bibfield{author}{\bibinfo{person}{Juan Miguel~Lopez Alcaraz} {and}
  \bibinfo{person}{Nils Strodthoff}.} \bibinfo{year}{2022}\natexlab{}.
\newblock \showarticletitle{Diffusion-based Time Series Imputation and
  Forecasting with Structured State Space Models}.
\newblock \bibinfo{journal}{\emph{arXiv preprint arXiv:2208.09399}}
  (\bibinfo{year}{2022}).
\newblock


\bibitem[Allen(2004)]%
        {allen2004monitoring}
\bibfield{author}{\bibinfo{person}{Bruce Allen}.}
  \bibinfo{year}{2004}\natexlab{}.
\newblock \showarticletitle{Monitoring hard disks with SMART}.
\newblock \bibinfo{journal}{\emph{Linux Journal}} \bibinfo{volume}{2004},
  \bibinfo{number}{117} (\bibinfo{year}{2004}), \bibinfo{pages}{9}.
\newblock


\bibitem[Atkinson et~al\mbox{.}(2011)]%
        {atkinson2011numerical}
\bibfield{author}{\bibinfo{person}{Kendall Atkinson}, \bibinfo{person}{Weimin
  Han}, {and} \bibinfo{person}{David~E Stewart}.}
  \bibinfo{year}{2011}\natexlab{}.
\newblock \bibinfo{booktitle}{\emph{Numerical solution of ordinary differential
  equations}}.
\newblock \bibinfo{publisher}{John Wiley \& Sons}.
\newblock


\bibitem[Botezatu et~al\mbox{.}(2016)]%
        {botezatu2016predicting}
\bibfield{author}{\bibinfo{person}{Mirela~Madalina Botezatu},
  \bibinfo{person}{Ioana Giurgiu}, \bibinfo{person}{Jasmina Bogojeska}, {and}
  \bibinfo{person}{Dorothea Wiesmann}.} \bibinfo{year}{2016}\natexlab{}.
\newblock \showarticletitle{Predicting disk replacement towards reliable data
  centers}. In \bibinfo{booktitle}{\emph{Proceedings of the 22nd ACM SIGKDD
  International Conference on Knowledge Discovery and Data Mining}}.
  \bibinfo{pages}{39--48}.
\newblock


\bibitem[Cao et~al\mbox{.}(2018)]%
        {cao2018brits}
\bibfield{author}{\bibinfo{person}{Wei Cao}, \bibinfo{person}{Dong Wang},
  \bibinfo{person}{Jian Li}, \bibinfo{person}{Hao Zhou}, \bibinfo{person}{Lei
  Li}, {and} \bibinfo{person}{Yitan Li}.} \bibinfo{year}{2018}\natexlab{}.
\newblock \showarticletitle{Brits: Bidirectional recurrent imputation for time
  series}.
\newblock \bibinfo{journal}{\emph{Advances in neural information processing
  systems}}  \bibinfo{volume}{31} (\bibinfo{year}{2018}).
\newblock


\bibitem[Chai et~al\mbox{.}(2020)]%
        {chai2020human}
\bibfield{author}{\bibinfo{person}{Chengliang Chai}, \bibinfo{person}{Lei Cao},
  \bibinfo{person}{Guoliang Li}, \bibinfo{person}{Jian Li},
  \bibinfo{person}{Yuyu Luo}, {and} \bibinfo{person}{Samuel Madden}.}
  \bibinfo{year}{2020}\natexlab{}.
\newblock \showarticletitle{Human-in-the-loop outlier detection}. In
  \bibinfo{booktitle}{\emph{Proceedings of the 2020 ACM SIGMOD International
  Conference on Management of Data}}. \bibinfo{pages}{19--33}.
\newblock


\bibitem[Che et~al\mbox{.}(2018)]%
        {che2018recurrent}
\bibfield{author}{\bibinfo{person}{Zhengping Che}, \bibinfo{person}{Sanjay
  Purushotham}, \bibinfo{person}{Kyunghyun Cho}, \bibinfo{person}{David
  Sontag}, {and} \bibinfo{person}{Yan Liu}.} \bibinfo{year}{2018}\natexlab{}.
\newblock \showarticletitle{Recurrent neural networks for multivariate time
  series with missing values}.
\newblock \bibinfo{journal}{\emph{Scientific reports}} \bibinfo{volume}{8},
  \bibinfo{number}{1} (\bibinfo{year}{2018}), \bibinfo{pages}{1--12}.
\newblock


\bibitem[Chen et~al\mbox{.}(2019)]%
        {chen2019outage}
\bibfield{author}{\bibinfo{person}{Yujun Chen}, \bibinfo{person}{Xian Yang},
  \bibinfo{person}{Qingwei Lin}, \bibinfo{person}{Hongyu Zhang},
  \bibinfo{person}{Feng Gao}, \bibinfo{person}{Zhangwei Xu},
  \bibinfo{person}{Yingnong Dang}, \bibinfo{person}{Dongmei Zhang},
  \bibinfo{person}{Hang Dong}, \bibinfo{person}{Yong Xu}, {et~al\mbox{.}}}
  \bibinfo{year}{2019}\natexlab{}.
\newblock \showarticletitle{Outage prediction and diagnosis for cloud service
  systems}. In \bibinfo{booktitle}{\emph{The World Wide Web Conference}}.
  \bibinfo{pages}{2659--2665}.
\newblock


\bibitem[Deb et~al\mbox{.}(2017)]%
        {deb2017aesop}
\bibfield{author}{\bibinfo{person}{Supratim Deb}, \bibinfo{person}{Zihui Ge},
  \bibinfo{person}{Sastry Isukapalli}, \bibinfo{person}{Sarat Puthenpura},
  \bibinfo{person}{Shobha Venkataraman}, \bibinfo{person}{He Yan}, {and}
  \bibinfo{person}{Jennifer Yates}.} \bibinfo{year}{2017}\natexlab{}.
\newblock \showarticletitle{Aesop: Automatic policy learning for predicting and
  mitigating network service impairments}. In
  \bibinfo{booktitle}{\emph{Proceedings of the 23rd ACM SIGKDD International
  Conference on Knowledge Discovery and Data Mining}}.
  \bibinfo{pages}{1783--1792}.
\newblock


\bibitem[Devlin et~al\mbox{.}(2018)]%
        {devlin2018bert}
\bibfield{author}{\bibinfo{person}{Jacob Devlin}, \bibinfo{person}{Ming-Wei
  Chang}, \bibinfo{person}{Kenton Lee}, {and} \bibinfo{person}{Kristina
  Toutanova}.} \bibinfo{year}{2018}\natexlab{}.
\newblock \showarticletitle{Bert: Pre-training of deep bidirectional
  transformers for language understanding}.
\newblock \bibinfo{journal}{\emph{arXiv preprint arXiv:1810.04805}}
  (\bibinfo{year}{2018}).
\newblock


\bibitem[Fletcher~Mercaldo and Blume(2020)]%
        {fletcher2020missing}
\bibfield{author}{\bibinfo{person}{Sarah Fletcher~Mercaldo} {and}
  \bibinfo{person}{Jeffrey~D Blume}.} \bibinfo{year}{2020}\natexlab{}.
\newblock \showarticletitle{Missing data and prediction: the pattern submodel}.
\newblock \bibinfo{journal}{\emph{Biostatistics}} \bibinfo{volume}{21},
  \bibinfo{number}{2} (\bibinfo{year}{2020}), \bibinfo{pages}{236--252}.
\newblock


\bibitem[Fortuin et~al\mbox{.}(2020)]%
        {fortuin2020gp}
\bibfield{author}{\bibinfo{person}{Vincent Fortuin}, \bibinfo{person}{Dmitry
  Baranchuk}, \bibinfo{person}{Gunnar R{\"a}tsch}, {and}
  \bibinfo{person}{Stephan Mandt}.} \bibinfo{year}{2020}\natexlab{}.
\newblock \showarticletitle{Gp-vae: Deep probabilistic time series imputation}.
  In \bibinfo{booktitle}{\emph{International conference on artificial
  intelligence and statistics}}. PMLR, \bibinfo{pages}{1651--1661}.
\newblock


\bibitem[Gao et~al\mbox{.}(2020)]%
        {gao2020task}
\bibfield{author}{\bibinfo{person}{Jiechao Gao}, \bibinfo{person}{Haoyu Wang},
  {and} \bibinfo{person}{Haiying Shen}.} \bibinfo{year}{2020}\natexlab{}.
\newblock \showarticletitle{Task failure prediction in cloud data centers using
  deep learning}.
\newblock \bibinfo{journal}{\emph{IEEE transactions on services computing}}
  (\bibinfo{year}{2020}).
\newblock


\bibitem[Ho et~al\mbox{.}(2020)]%
        {ho2020denoising}
\bibfield{author}{\bibinfo{person}{Jonathan Ho}, \bibinfo{person}{Ajay Jain},
  {and} \bibinfo{person}{Pieter Abbeel}.} \bibinfo{year}{2020}\natexlab{}.
\newblock \showarticletitle{Denoising diffusion probabilistic models}.
\newblock \bibinfo{journal}{\emph{Advances in Neural Information Processing
  Systems}}  \bibinfo{volume}{33} (\bibinfo{year}{2020}),
  \bibinfo{pages}{6840--6851}.
\newblock


\bibitem[Jayathilaka et~al\mbox{.}(2017)]%
        {jayathilaka2017performance}
\bibfield{author}{\bibinfo{person}{Hiranya Jayathilaka},
  \bibinfo{person}{Chandra Krintz}, {and} \bibinfo{person}{Rich Wolski}.}
  \bibinfo{year}{2017}\natexlab{}.
\newblock \showarticletitle{Performance monitoring and root cause analysis for
  cloud-hosted web applications}. In \bibinfo{booktitle}{\emph{Proceedings of
  the 26th International Conference on World Wide Web}}.
  \bibinfo{pages}{469--478}.
\newblock


\bibitem[Jolicoeur-Martineau et~al\mbox{.}(2021)]%
        {jolicoeur2021gotta}
\bibfield{author}{\bibinfo{person}{Alexia Jolicoeur-Martineau},
  \bibinfo{person}{Ke Li}, \bibinfo{person}{R{\'e}mi Pich{\'e}-Taillefer},
  \bibinfo{person}{Tal Kachman}, {and} \bibinfo{person}{Ioannis Mitliagkas}.}
  \bibinfo{year}{2021}\natexlab{}.
\newblock \showarticletitle{Gotta go fast when generating data with score-based
  models}.
\newblock \bibinfo{journal}{\emph{arXiv preprint arXiv:2105.14080}}
  (\bibinfo{year}{2021}).
\newblock


\bibitem[Kingma and Welling(2013)]%
        {kingma2013auto}
\bibfield{author}{\bibinfo{person}{Diederik~P Kingma} {and}
  \bibinfo{person}{Max Welling}.} \bibinfo{year}{2013}\natexlab{}.
\newblock \showarticletitle{Auto-encoding variational bayes}.
\newblock \bibinfo{journal}{\emph{arXiv preprint arXiv:1312.6114}}
  (\bibinfo{year}{2013}).
\newblock


\bibitem[Levy et~al\mbox{.}(2020)]%
        {levy2020predictive}
\bibfield{author}{\bibinfo{person}{Sebastien Levy}, \bibinfo{person}{Randolph
  Yao}, \bibinfo{person}{Youjiang Wu}, \bibinfo{person}{Yingnong Dang},
  \bibinfo{person}{Peng Huang}, \bibinfo{person}{Zheng Mu}, \bibinfo{person}{Pu
  Zhao}, \bibinfo{person}{Tarun Ramani}, \bibinfo{person}{Naga Govindaraju},
  \bibinfo{person}{Xukun Li}, {et~al\mbox{.}}} \bibinfo{year}{2020}\natexlab{}.
\newblock \showarticletitle{Predictive and Adaptive Failure Mitigation to Avert
  Production Cloud $\{$VM$\}$ Interruptions}. In \bibinfo{booktitle}{\emph{14th
  USENIX Symposium on Operating Systems Design and Implementation (OSDI 20)}}.
  \bibinfo{pages}{1155--1170}.
\newblock


\bibitem[Little and Rubin(2002)]%
        {little2002single}
\bibfield{author}{\bibinfo{person}{Roderick~JA Little} {and}
  \bibinfo{person}{Donald~B Rubin}.} \bibinfo{year}{2002}\natexlab{}.
\newblock \showarticletitle{Single imputation methods}.
\newblock \bibinfo{journal}{\emph{Statistical analysis with missing data}}
  (\bibinfo{year}{2002}), \bibinfo{pages}{59--74}.
\newblock


\bibitem[Little et~al\mbox{.}(2014)]%
        {little2014joys}
\bibfield{author}{\bibinfo{person}{Todd~D Little}, \bibinfo{person}{Terrence~D
  Jorgensen}, \bibinfo{person}{Kyle~M Lang}, {and} \bibinfo{person}{E~Whitney~G
  Moore}.} \bibinfo{year}{2014}\natexlab{}.
\newblock \showarticletitle{On the joys of missing data}.
\newblock \bibinfo{journal}{\emph{Journal of pediatric psychology}}
  \bibinfo{volume}{39}, \bibinfo{number}{2} (\bibinfo{year}{2014}),
  \bibinfo{pages}{151--162}.
\newblock


\bibitem[Liu et~al\mbox{.}(2022)]%
        {liu2022multi}
\bibfield{author}{\bibinfo{person}{Yudong Liu}, \bibinfo{person}{Hailan Yang},
  \bibinfo{person}{Pu Zhao}, \bibinfo{person}{Minghua Ma},
  \bibinfo{person}{Chengwu Wen}, \bibinfo{person}{Hongyu Zhang},
  \bibinfo{person}{Chuan Luo}, \bibinfo{person}{Qingwei Lin},
  \bibinfo{person}{Chang Yi}, \bibinfo{person}{Jiaojian Wang}, {et~al\mbox{.}}}
  \bibinfo{year}{2022}\natexlab{}.
\newblock \showarticletitle{Multi-task Hierarchical Classification for Disk
  Failure Prediction in Online Service Systems}. In
  \bibinfo{booktitle}{\emph{Proceedings of the 28th ACM SIGKDD Conference on
  Knowledge Discovery and Data Mining}}. \bibinfo{pages}{3438--3446}.
\newblock


\bibitem[Liu et~al\mbox{.}(2019)]%
        {liu2019naomi}
\bibfield{author}{\bibinfo{person}{Yukai Liu}, \bibinfo{person}{Rose Yu},
  \bibinfo{person}{Stephan Zheng}, \bibinfo{person}{Eric Zhan}, {and}
  \bibinfo{person}{Yisong Yue}.} \bibinfo{year}{2019}\natexlab{}.
\newblock \showarticletitle{Naomi: Non-autoregressive multiresolution sequence
  imputation}.
\newblock \bibinfo{journal}{\emph{Advances in neural information processing
  systems}}  \bibinfo{volume}{32} (\bibinfo{year}{2019}).
\newblock


\bibitem[Lu et~al\mbox{.}(2022)]%
        {lu2022dpm}
\bibfield{author}{\bibinfo{person}{Cheng Lu}, \bibinfo{person}{Yuhao Zhou},
  \bibinfo{person}{Fan Bao}, \bibinfo{person}{Jianfei Chen},
  \bibinfo{person}{Chongxuan Li}, {and} \bibinfo{person}{Jun Zhu}.}
  \bibinfo{year}{2022}\natexlab{}.
\newblock \showarticletitle{DPM-Solver: A Fast ODE Solver for Diffusion
  Probabilistic Model Sampling in Around 10 Steps}.
\newblock \bibinfo{journal}{\emph{arXiv preprint arXiv:2206.00927}}
  (\bibinfo{year}{2022}).
\newblock


\bibitem[Lu et~al\mbox{.}(2020)]%
        {lu2020making}
\bibfield{author}{\bibinfo{person}{Sidi Lu}, \bibinfo{person}{Bing Luo},
  \bibinfo{person}{Tirthak Patel}, \bibinfo{person}{Yongtao Yao},
  \bibinfo{person}{Devesh Tiwari}, {and} \bibinfo{person}{Weisong Shi}.}
  \bibinfo{year}{2020}\natexlab{}.
\newblock \showarticletitle{Making Disk Failure Predictions $\{$SMARTer$\}$!}.
  In \bibinfo{booktitle}{\emph{18th USENIX Conference on File and Storage
  Technologies (FAST 20)}}. \bibinfo{pages}{151--167}.
\newblock


\bibitem[Luo et~al\mbox{.}(2021)]%
        {luo2021ntam}
\bibfield{author}{\bibinfo{person}{Chuan Luo}, \bibinfo{person}{Pu Zhao},
  \bibinfo{person}{Bo Qiao}, \bibinfo{person}{Youjiang Wu},
  \bibinfo{person}{Hongyu Zhang}, \bibinfo{person}{Wei Wu},
  \bibinfo{person}{Weihai Lu}, \bibinfo{person}{Yingnong Dang},
  \bibinfo{person}{Saravanakumar Rajmohan}, \bibinfo{person}{Qingwei Lin},
  {et~al\mbox{.}}} \bibinfo{year}{2021}\natexlab{}.
\newblock \showarticletitle{NTAM: neighborhood-temporal attention model for
  disk failure prediction in cloud platforms}. In
  \bibinfo{booktitle}{\emph{Proceedings of the Web Conference 2021}}.
  \bibinfo{pages}{1181--1191}.
\newblock


\bibitem[Luo et~al\mbox{.}(2018)]%
        {luo2018multivariate}
\bibfield{author}{\bibinfo{person}{Yonghong Luo}, \bibinfo{person}{Xiangrui
  Cai}, \bibinfo{person}{Ying Zhang}, \bibinfo{person}{Jun Xu},
  {et~al\mbox{.}}} \bibinfo{year}{2018}\natexlab{}.
\newblock \showarticletitle{Multivariate time series imputation with generative
  adversarial networks}.
\newblock \bibinfo{journal}{\emph{Advances in neural information processing
  systems}}  \bibinfo{volume}{31} (\bibinfo{year}{2018}).
\newblock


\bibitem[Ma et~al\mbox{.}(2022)]%
        {ma2022empirical}
\bibfield{author}{\bibinfo{person}{Minghua Ma}, \bibinfo{person}{Yudong Liu},
  \bibinfo{person}{Yuang Tong}, \bibinfo{person}{Haozhe Li},
  \bibinfo{person}{Pu Zhao}, \bibinfo{person}{Yong Xu}, \bibinfo{person}{Hongyu
  Zhang}, \bibinfo{person}{Shilin He}, \bibinfo{person}{Lu Wang},
  \bibinfo{person}{Yingnong Dang}, \bibinfo{person}{Saravanakumar Rajmohan},
  {and} \bibinfo{person}{Qingwei Lin}.} \bibinfo{year}{2022}\natexlab{}.
\newblock \showarticletitle{An Empirical Investigation of Missing Data Handling
  in Cloud Node Failure Prediction}. In \bibinfo{booktitle}{\emph{Proceedings
  of the European Software Engineering Conference and Symposium on the
  Foundations of Software Engineering (ESEC/FSE)}}. \bibinfo{pages}{1453 --
  1464}.
\newblock


\bibitem[Matheson and Winkler(1976)]%
        {matheson1976scoring}
\bibfield{author}{\bibinfo{person}{James~E Matheson} {and}
  \bibinfo{person}{Robert~L Winkler}.} \bibinfo{year}{1976}\natexlab{}.
\newblock \showarticletitle{Scoring rules for continuous probability
  distributions}.
\newblock \bibinfo{journal}{\emph{Management science}} \bibinfo{volume}{22},
  \bibinfo{number}{10} (\bibinfo{year}{1976}), \bibinfo{pages}{1087--1096}.
\newblock


\bibitem[Meza et~al\mbox{.}(2015)]%
        {meza2015large}
\bibfield{author}{\bibinfo{person}{Justin Meza}, \bibinfo{person}{Qiang Wu},
  \bibinfo{person}{Sanjev Kumar}, {and} \bibinfo{person}{Onur Mutlu}.}
  \bibinfo{year}{2015}\natexlab{}.
\newblock \showarticletitle{A large-scale study of flash memory failures in the
  field}.
\newblock \bibinfo{journal}{\emph{ACM SIGMETRICS Performance Evaluation
  Review}} \bibinfo{volume}{43}, \bibinfo{number}{1} (\bibinfo{year}{2015}),
  \bibinfo{pages}{177--190}.
\newblock


\bibitem[Pratama et~al\mbox{.}(2016)]%
        {pratama2016review}
\bibfield{author}{\bibinfo{person}{Irfan Pratama},
  \bibinfo{person}{Adhistya~Erna Permanasari}, \bibinfo{person}{Igi Ardiyanto},
  {and} \bibinfo{person}{Rini Indrayani}.} \bibinfo{year}{2016}\natexlab{}.
\newblock \showarticletitle{A review of missing values handling methods on
  time-series data}. In \bibinfo{booktitle}{\emph{2016 international conference
  on information technology systems and innovation (ICITSI)}}. IEEE,
  \bibinfo{pages}{1--6}.
\newblock


\bibitem[Read(1999)]%
        {read1999linear}
\bibfield{author}{\bibinfo{person}{ALEXANDER~L Read}.}
  \bibinfo{year}{1999}\natexlab{}.
\newblock \showarticletitle{Linear interpolation of histograms}.
\newblock \bibinfo{journal}{\emph{Nuclear Instruments and Methods in Physics
  Research Section A: Accelerators, Spectrometers, Detectors and Associated
  Equipment}} \bibinfo{volume}{425}, \bibinfo{number}{1-2}
  (\bibinfo{year}{1999}), \bibinfo{pages}{357--360}.
\newblock


\bibitem[Sankar et~al\mbox{.}(2013)]%
        {sankar2013datacenter}
\bibfield{author}{\bibinfo{person}{Sriram Sankar}, \bibinfo{person}{Mark Shaw},
  \bibinfo{person}{Kushagra Vaid}, {and} \bibinfo{person}{Sudhanva
  Gurumurthi}.} \bibinfo{year}{2013}\natexlab{}.
\newblock \showarticletitle{Datacenter scale evaluation of the impact of
  temperature on hard disk drive failures}.
\newblock \bibinfo{journal}{\emph{ACM Transactions on Storage (TOS)}}
  \bibinfo{volume}{9}, \bibinfo{number}{2} (\bibinfo{year}{2013}),
  \bibinfo{pages}{1--24}.
\newblock


\bibitem[Sohl-Dickstein et~al\mbox{.}(2015)]%
        {sohl2015deep}
\bibfield{author}{\bibinfo{person}{Jascha Sohl-Dickstein},
  \bibinfo{person}{Eric Weiss}, \bibinfo{person}{Niru Maheswaranathan}, {and}
  \bibinfo{person}{Surya Ganguli}.} \bibinfo{year}{2015}\natexlab{}.
\newblock \showarticletitle{Deep unsupervised learning using nonequilibrium
  thermodynamics}. In \bibinfo{booktitle}{\emph{International Conference on
  Machine Learning}}. PMLR, \bibinfo{pages}{2256--2265}.
\newblock


\bibitem[Song et~al\mbox{.}(2020)]%
        {song2020score}
\bibfield{author}{\bibinfo{person}{Yang Song}, \bibinfo{person}{Jascha
  Sohl-Dickstein}, \bibinfo{person}{Diederik~P Kingma},
  \bibinfo{person}{Abhishek Kumar}, \bibinfo{person}{Stefano Ermon}, {and}
  \bibinfo{person}{Ben Poole}.} \bibinfo{year}{2020}\natexlab{}.
\newblock \showarticletitle{Score-based generative modeling through stochastic
  differential equations}.
\newblock \bibinfo{journal}{\emph{arXiv preprint arXiv:2011.13456}}
  (\bibinfo{year}{2020}).
\newblock


\bibitem[Sun et~al\mbox{.}(2019)]%
        {sun2019system}
\bibfield{author}{\bibinfo{person}{Xiaoyi Sun}, \bibinfo{person}{Krishnendu
  Chakrabarty}, \bibinfo{person}{Ruirui Huang}, \bibinfo{person}{Yiquan Chen},
  \bibinfo{person}{Bing Zhao}, \bibinfo{person}{Hai Cao},
  \bibinfo{person}{Yinhe Han}, \bibinfo{person}{Xiaoyao Liang}, {and}
  \bibinfo{person}{Li Jiang}.} \bibinfo{year}{2019}\natexlab{}.
\newblock \showarticletitle{System-level hardware failure prediction using deep
  learning}. In \bibinfo{booktitle}{\emph{2019 56th ACM/IEEE design automation
  conference (DAC)}}. IEEE, \bibinfo{pages}{1--6}.
\newblock


\bibitem[Tashiro et~al\mbox{.}(2021)]%
        {tashiro2021csdi}
\bibfield{author}{\bibinfo{person}{Yusuke Tashiro}, \bibinfo{person}{Jiaming
  Song}, \bibinfo{person}{Yang Song}, {and} \bibinfo{person}{Stefano Ermon}.}
  \bibinfo{year}{2021}\natexlab{}.
\newblock \showarticletitle{CSDI: Conditional score-based diffusion models for
  probabilistic time series imputation}.
\newblock \bibinfo{journal}{\emph{Advances in Neural Information Processing
  Systems}}  \bibinfo{volume}{34} (\bibinfo{year}{2021}),
  \bibinfo{pages}{24804--24816}.
\newblock


\bibitem[Ward and Barker(2014)]%
        {ward2014observing}
\bibfield{author}{\bibinfo{person}{Jonathan~Stuart Ward} {and}
  \bibinfo{person}{Adam Barker}.} \bibinfo{year}{2014}\natexlab{}.
\newblock \showarticletitle{Observing the clouds: a survey and taxonomy of
  cloud monitoring}.
\newblock \bibinfo{journal}{\emph{Journal of Cloud Computing}}
  \bibinfo{volume}{3}, \bibinfo{number}{1} (\bibinfo{year}{2014}),
  \bibinfo{pages}{1--30}.
\newblock


\bibitem[Zhang et~al\mbox{.}(2018)]%
        {zhang2018layerwise}
\bibfield{author}{\bibinfo{person}{Jianguo Zhang}, \bibinfo{person}{Ji Wang},
  \bibinfo{person}{Lifang He}, \bibinfo{person}{Zhao Li}, {and}
  \bibinfo{person}{S~Yu Philip}.} \bibinfo{year}{2018}\natexlab{}.
\newblock \showarticletitle{Layerwise perturbation-based adversarial training
  for hard drive health degree prediction}. In \bibinfo{booktitle}{\emph{2018
  IEEE International Conference on Data Mining (ICDM)}}. IEEE,
  \bibinfo{pages}{1428--1433}.
\newblock


\end{thebibliography}

%%
%% If your work has an appendix, this is the place to put it.
\appendix

\end{document}